# High temperature thermal cycling effect on the irreversible responses of lattice structure, magnetic properties and electrical conductivity in $Co_{2.75}Fe_{0.25}O_{4+\delta}$ spinel oxide

Rabindra Nath Bhowmik[*a], Peram D Babu[b], Anil Kumar Sinha[c,d], and Abhay Bhisikar[c]

[a]Department of Physics, Pondicherry University, R.V. Nagar, Kalapet, Pondicherry 605014, India

[b]UGC-DAE Consortium for Scientific Research, Mumbai Centre, Bhabha Atomic Research Centre, Trombay, Mumbai-400085, India

[c]HXAL, SUS, Raja Ramanna Centre for Advanced Technology, Indore- 452013, India

[d]Homi Bhabha National Institute, Anushakti Nagar, Mumbai -400 094 India

[*]Corresponding author: Tel.: +91-9944064547; E-mail: rnbhowmik.phy@pondiuni.edu.in

**ABSTRACT**:

We report high temperature synchrotron X-ray diffraction (SXRD), magnetization and current-voltage (I-V) characteristics for the samples of $Co_{2.75}Fe_{0.25}O_4$ ferrite. The material was prepared by chemical reaction of the Fe and Co nitrate solutions at pH ~ 11 and subsequent thermal annealing. Physical properties of the samples were measured by cycling the temperature from 300 K to high temperature (warming mode) and return back to 300 K (cooling mode). The lattice structure showed sensitivity to high measurement temperatures. Magnetization curves showed defect induced ferromagnetic phase at higher temperatures and superparamagnetic blocking of the ferrimagnetic particles near to 300 K or below. Electrical conductivity exhibited thermal hysteresis loop at higher measurement temperatures. The samples exhibited new form (not studied so far) of surface magnetism in Co rich spinel oxides and irreversibility phenomena in lattice structure, magnetization and conductivity on cycling the measurement temperatures.

■ **INTRODUCTION**

The spinel oxides are strongly spin correlated electronic system. They are defined by formula unit $AB_2O_4$, where metal ions occupy two inequivalent lattice sites A and B with tetrahedrally and octahedrally coordinated oxygen ions, respectively in cubic crystal with space group $Fd\overline{3}m$ [1-2]. The distribution of Co and Fe ions at A and B sites determine magnetic and electrical properties in spinel structure. Co rich spinel oxides ($Fe_{3-x}Co_xO_4$; $1<x<3$) are interesting for their ability to tune properties by altering the synthesis route, heat treatment, and Co content in the equilibrium and non-equilibrium cubic spinel structure [3-11]. Structural phase of Co rich spinels that lie in the miscibility gap of the $Fe_3O_4$-$Co_3O_4$ phase diagram is kinetically sensitive to Co content [1-3, 9, 11]. The annealing in the temperature range 900–950 $^0$C form stable phase. Otherwise, cubic spinel phase of Co rich spinel ferrites shows spinodal decomposition (Co-rich and Fe-rich phases). The spinodal decomposition is also formed in thin films of $Co_{1.8}Fe_{1.2}O_4$ (Co rich and Fe rich phases) [12] and Mn ferrite (Mn rich and Fe rich phases) [13]. The self-composite material is new for studying heterogeneoud chemical structure in oxides, which is characteristically different from hybrid composite of two different compounds [14].

We cite some of the works that reported high temperature structure and properties of spinel pxides. For example, $CuFe_2O_4$ spinel oxide showed a fully inverted tetragonal phase in the temperature range 2 K to 600 K and a cubic spinel phase at 660 K [15]. Both the phases coexisted up to 700 K and cubic phase alone is stabilized at higher temperatures. Magnetically, ferrimagnetic structure of $CuFe_2O_4$ disappeared at $T_N$ ~750 K. The $Mn_xFe_{3-x}O_4$ oxide at room temperature formed cubic phase for $x \leq 1$ and tetragonal distortion for $1 < x \leq 3$ [16]. FeMn$_2$O$_4$ showed a structural phase transition at $\sim$ 595 K from (low temperature) tetragonal phase to (high temperature) cubic phase with ferrimagnetic transitions at $\sim$ 373 K and $\sim$ 50 K [17]. O'Neill et al. [18] performed neutron diffraction to study the thermal cycling effect on cation order-disorder in Mg$_2$TiO$_4$. The distribution of Mg$^{2+}$ and Ti$^{4+}$ ions was found in ordered state (inverse structure with all Ti$^{4+}$ ions occupy half of the B sites) when temperature warmed from 300 K to 1173 K and a disordered state (migration of a fraction of Ti$^{4+}$ to A sites) at higher temperatures. The rate of reordering of the cations between A and B sites was faster during cooling process and high-temperature disordered state is not preserved after cooling back to 300 K. The high temperature XRD patterns suggested a correlation between cell parameter and distribution of cations in MgFe$_2$O$_4$ spinel oxide during thermal cycling (warming $\leftrightarrow$ cooling) of measurements [19].

The intrinsic point defects, introduced during thermal cycling, generally perturb the structure of cations from their equilibrium state and modify the properties [20-22]. Despite of showing many unusual low temperature magnetic properties in Co rich spinel oxides [2, 3, 6, 9, 11, 23], the high temperature study is limited [10]. In this work, we studied high temperature properties (structure, magnetism and electrical conductivity) for a highly Co rich spinel oxide Co$_{2.75}$Fe$_{0.25}$O$_4$ and discussed the role of thermal cycle induced irreversible effects on properties. We performed the measurements in air to realize the naturally expected defects in spinel oxides. The materials with inhomogeneous lattice structure, magnetic spin order and electronic structure are expected to be interesting for studying first order magnetic phase transition, and achieving high magneto-caloric effect, high catalytic and surface chemical kinetics [11, 24-26].

■ **EXPERIMENTAL SECTION**

**Sample preparation.** The spinel oxide $Co_{2.75}Fe_{0.25}O_4$ was prepared through chemical reaction of the required amounts of $Co(NO_3)_2.6H_2O$ and $Fe(No_3)_3.9H_2O$ salts. The salts were dissolved in distilled water to yield a transparent solution at pH $\sim$1.4. NaOH solution with initial pH $\sim$13 was used as the precipitating agent and added into the nitrate solution until pH reached to $\sim$ 11. The reaction temperature was maintained at 80 °C for 4h with continuous stirring. The pH was maintained at 11 by adding required amount of NaOH solution during chemical reaction. The products were allowed to cool down to room temperature and allowed to precipitate at the bottom of a Borosil beaker. The transparent solution from the top portion was removed carefully and remaining product was washed several times with distilled water and each time it was dried at 100°C. Finally, the resultant powder was heated at 200-250 °C in a beaker to confirm the complete removal of the bi-product of $NaNO_3$, which formed a white coating on the wall of beaker and black coloured (magnetic) powder was collected at the centre of the beaker when placed on a Rotamantle. The collected black powder was made into several pellets and annealed at selected temperatures in the range 200-900$^o$C for 6h in air. The prepared samples were denoted as CF_20, CF_50and CF_90 for annealing temperature at 200 $^0$C, 500$^0$C and 900 $^0$C, respectively. The heating and cooling rate was maintained @ 5 °C/min.

**Sample characterization.** The synchrotron X-ray diffraction (SXRD) patterns were recorded in the 2θ range 5-40 ° for samples CF_20 and CF_90 using synchrotron radiation facilities at angle dispersive X-ray diffraction beam line 12 of Indus-2, Indore, India. The wave length was fixed at 0.7820 Å and 0.7600 Å for the samples CF_20 and CF_90, respectively. Photon energy and the sample to detector distance for SXRD were calibrated by using SXRD pattern of $LaB_6$ NIST standard. The high temperature SXRD measurements were performed in the temperature range (300 K to 873 K) in air with temperature variation @ 20 °C/min and temperature stabilization for nearly 2 minutes before measurement at each temperature. The dc magnetization [M(T)] in the warming mode (300 K-950 K) and cooling mode (back to 300 K) were measured using physical properties measurement system (PPMS-EC2, Quantum Design, USA). The field dependent magnetization [M(H)] data were recorded within magnetic field range ± 70 kOe at selected temperatures in the range 300 K-900 K during warming mode after completing the M(T) measurement cycle. The current-voltage (I-V) curves for CF_50 and CF_90 samples were measured in the temperature range 300-623 K using Keithley 6517B high resistance meter. The disc shaped samples ($\varnothing$= 10 mm, t~0.5 mm) were placed between Pt electrodes of a home-made sample holder to make a Pt/sample/Pt device structure. The current passing through the sample was measured by sweeping the dc voltage within ± 50 V at each measurement temperature.

■ **RESULTS**

**Structural Properties.** Figure 1 shows SXRD pattern of the CF_20 and CF_90 samples at selected temperatures. The Y-axis has been re-scaled for some patterns to compare between warming and cooling modes. SXRD pattern at all measurement temperatures (300-873 K) show

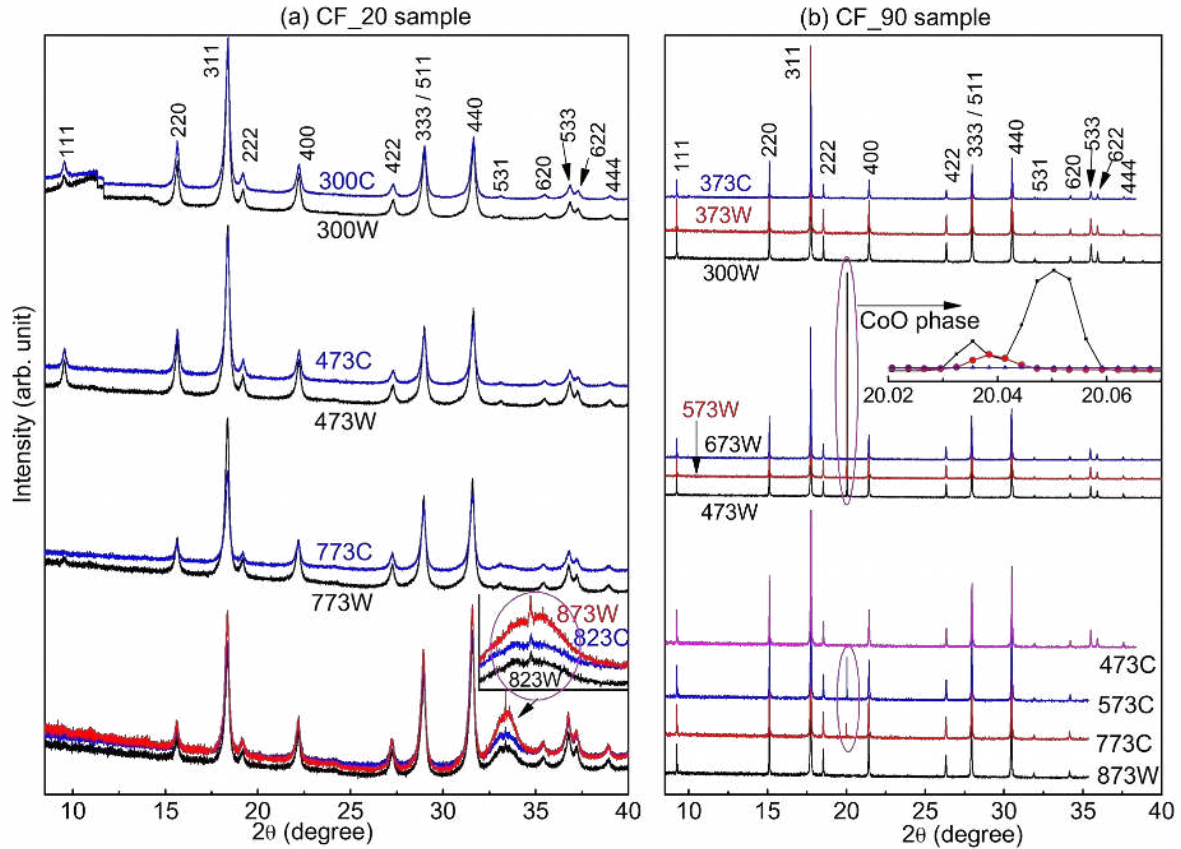

Figure 1. Synchrotron XRD pattern measured in the temperature range 300-873 K during warming and cooling modes for the CF_20 (a) and CF_90 (b) samples. Inset shows additional disorder/phase at higher temperatures.

cubic spinel structure (space group Fd$\bar{3}$m) for both the samples [5, 7]. The crystalline planes corresponding to cubic spinel structure are indexed at the top of Figure 1(a-b). In CF_20 sample, the usual peak intensity ratio is maintained for temperatures up to 773 K. At higher temperatures (825K and 873K), a sharp peak same as (531) plane sits on the amorphous background at 2θ range 32.5-34.5º (inset of Figure 1(a)). The background introduces a butterfly type wing; an indication of structural distortion. Subsequently, a preferential orientation of the (511) and (440) planes showed unusual increase of intensity in contrast to usually observed highest intensity for (311) plane. The SXRD patterns possibility of any impurity phases, such as α-Fe$_2$O$_3$ and CoO [2, 4, 8]. But, a non-equilibrium lattice structure is expected due to low annealing temperature (200 $^0$

C) [5, 27]. On the other hand, CFO_90 sample (Figure 1(b)) showed sharp crystalline peaks. The cubic spinel structure is maintained at 300 K and 373 K during warming mode. In addition to cubic spinel phase, an exceptionally high intensity (200) peak of CoO phase (cubic structure with space group Pm3m) [4-5] is seen at $2\theta \sim 20^{0}$ for temperature at 473 K (inset of Figure 1(b)). The CoO phase is grain oriented along (200) direction. Intensity of the CoO phase decreased at 573 K and finally disappeared at $\geq$ 673 K during warming mode. The CoO phase reappeared during cooling mode with less intensity in the range 773-573 K and is absent in the temperature range 473-300 K. A close scrutiny of the shape and asymmetric nature of the peaks indicates a possibility of two-phase cubic spinel structure. Figure 2 shows Rietveld refinement of the SXRD

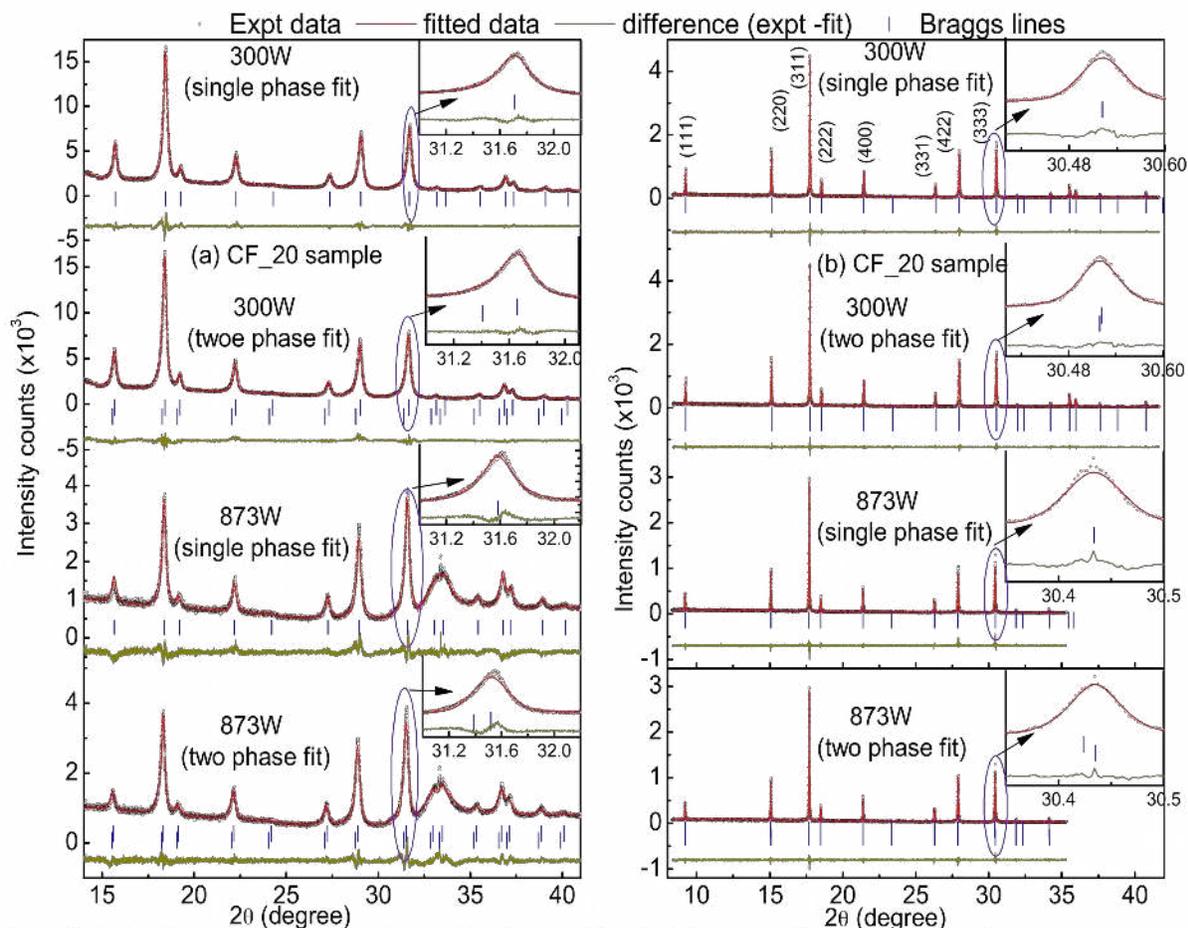

Figure 2. Rietveld refined patterns of the CF_20 (a) and CF_90 (b) samples fitted with single phase and two phases for the experimental data at 300W and 873W. The insets clarify the single phased and two-phased fit for (333) peak.

patterns at 300 K and 873 K of the CF_20 and CF_90 samples in warming mode. Details of the refinement process are included in supplementary information. The SXRD patterns have been fitted using single phase and two phase models of spinel structure. The single phase is modeled by full occupancy of 8a sites for $Co^{2+}$ ions and 16 sites are co-occupied by $Co^{3+}$ and $Fe^{3+}$ ions. The difference in the fit with single phase and two phase components is shown in the insets of Figure 2 for (333) peak. The two-phased model is best suited for CF_20 sample. The difference between two phase components of spinel structure is negligible for CF_90 sample. The peak of CoO phase in CF_90 sample was excluded during fit of cubic spinel phase. The two phase model for CF_20 sample is fitted by assuming Co rich phase (phase 1), where 8a sites are fully occupied by Co ions and Fe content at 16d sites is assigned less amount than the assigned value for single phase, and Fe rich phase (phase 2), where Co and Fe are allowed to co-occupy both at 8a and 16d sites and the total Fe content is more than the assigned value for single phase model. Table 1 provides parameters obtained from fitting of SXRD data at 300 K using two-phase cubic spinel structure for CF_20 and single phase model for CF_90 samples at 300 K. Rest of the fit parameters at 300 K and 873 K are shown in Table S1 (supplementary). We found that oxygen content and composition of metal ions nearly matched to chemical composition of the samples at 300 K, where as composition at 873 K shows excess O atoms per formula unit of the spinel structure. The fraction of spinel phase components for CF_20 sample has been calculated from structural refinement (consistent to average value of the calculation from intensity ratio using (311), (400), (333) peaks). The fraction of spinel phase and CoO phase in CF_90 sample has been calculated from intensity ratio of the (311) peak for spinel phase and (200) peak for CoO. Figure 3 schematically shows temperature variation of the fraction of phases for CF_20 and CF_90 samples during warming and cooling modes. The Co-rich phase dominates than Fe-rich

Table 1. Structural parameters from Rietveld refinement parameters of SXRD data at 300 K (warming mode)

(a) single phase model of cubic spinel structure for CF_20 sample and CF_90 sample

| Atoms (sites) | CF_20 sample | | | | | CF_90 sample | | | | |
|---|---|---|---|---|---|---|---|---|---|---|
| | Wyckoff positions | | | B | Occupancy | Wyckoff positions | | | B | Occupancy |
| | X | Y | Z | | | X | Y | Z | | |
| Co(8a) | 0.12500 | 0.12500 | 0.12500 | 0.97 | 1.000 | 0.12500 | 0.12500 | 0.12500 | 0.97 | 1.000 |
| Fe(8a) | 0.12500 | 0.12500 | 0.12500 | 0.97 | 0.000 | 0.12500 | 0.12500 | 0.12500 | 0.97 | 0.000 |
| Fe(16d) | 0.50000 | 0.50000 | 0.50000 | 0.97 | 0.250 | 0.50000 | 0.50000 | 0.50000 | 0.97 | 0.250 |
| Co(16d) | 0.50000 | 0.50000 | 0.50000 | 0.97 | 1.750 | 0.50000 | 0.50000 | 0.50000 | 0.97 | 1.750 |
| O (32e) | 0.26020(33) | 0.26020(33) | 0.26020(33) | 0.78 | 3.962(29) | 0.26020(33) | 0.26020(33) | 0.26020(33) | 0.78 | 4.049(29) |
| Cell parameter s: $a = 8.10075$ (33) Å, $V = 531.588$ (37) Å³ | | | | | | Cell parameters: $a = 8.16673(12)$ Å, $V = 544.683(14)$ Å³ | | | | |
| $R_p$: 3.52, $R_{wp}$: 4.54, $R_{exp}$: 2.28, $\chi^2$: 3.96 | | | | | | $R_p$: 9.44, $R_{wp}$: 12.4, $R_{exp}$: 10.94, $\chi^2$: 1.28 | | | | |

b) Two-phase model of cubic spinel structure for CF_20 sample

| Atoms (sites) | Co- rich phase | | | | Fe- rich phase | | | |
|---|---|---|---|---|---|---|---|---|
| | Wyckoff positions | | B | Occupancy | Wyckoff positions | | B | Occupancy |

|  | X | Y | Z |  |  | X | Y | Z |  |  |
|---|---|---|---|---|---|---|---|---|---|---|
| Co(8a) | 0.12500 | 0.12500 | 0.12500 | 0.97 | 1.000 | 0.12500 | 0.12500 | 0.12500 | 0.97 | 0.850 |
| Fe(8a) | 0.12500 | 0.12500 | 0.12500 | 0.97 | 0.000 | 0.12500 | 0.12500 | 0.12500 | 0.97 | 0.150 |
| Fe(16d) | 0.50000 | 0.50000 | 0.50000 | 0.79 | 0.050 | 0.50000 | 0.50000 | 0.50000 | 0.79 | 0.190 |
| Co(16d) | 0.50000 | 0.50000 | 0.50000 | 0.79 | 1.950(0) | 0.50000 | 0.50000 | 0.50000 | 0.79 | 1.750 |
| O (32e) | 0.26119(28) | 0.26119(28) | 0.26119(28) | 0.72 | 3.947(23) | 0.26250(48) | 0.26250(48) | 0.26250(48) | 0.72 | 4.640(54) |

Cell parameters: $a$ = 8.10992 (30) Å, $V$ = 533.396(34) Å³ | Cell parameters: $a$ = 8.17176 (71) Å, $V$ =545.692 (82)Å³

Co rich phase (Bragg R-factor: 4.57, phase fraction: 65.69%) | $R_p$: 4.04, $R_{wp}$: 5.08, $R_{exp}$: 2.39, $\chi^2$: 4.51

Fe rich phase (Bragg R-factor: 4.95, phase fraction: 34.31%)

spinel phase in CF_20 sample. The difference between Co-rich and Fe-rich phases decreased during cooling mode. This gives rise to a loop, indicating the gain of Fe rich phase by costing the Co-rich phase. A thermal hysteresis in the formation of CoO phase or intermediate kinetic instability in spinel structure is observed for CF_90 sample, where CoO phase (maximum ~ 80% at 473 K in warming mode) has reduced during cooling mode (maximum ~ 17 % at 573 K). Figure 4(a-f) shows the temperature variation of structural parameters (lattice constant ($a$), oxygen parameter ($u$) and $\delta$O (excess oxygen)) obtained using single phase model. All these parameters showed irreversible paths during thermal cycling of the measurement, irrespective of CF_20 (two spinel phase structure) or CF_90 (single spinel phase + intermediate CoO phase) samples. The lattice parameter shows usual thermal expansion, but higher values of the lattice

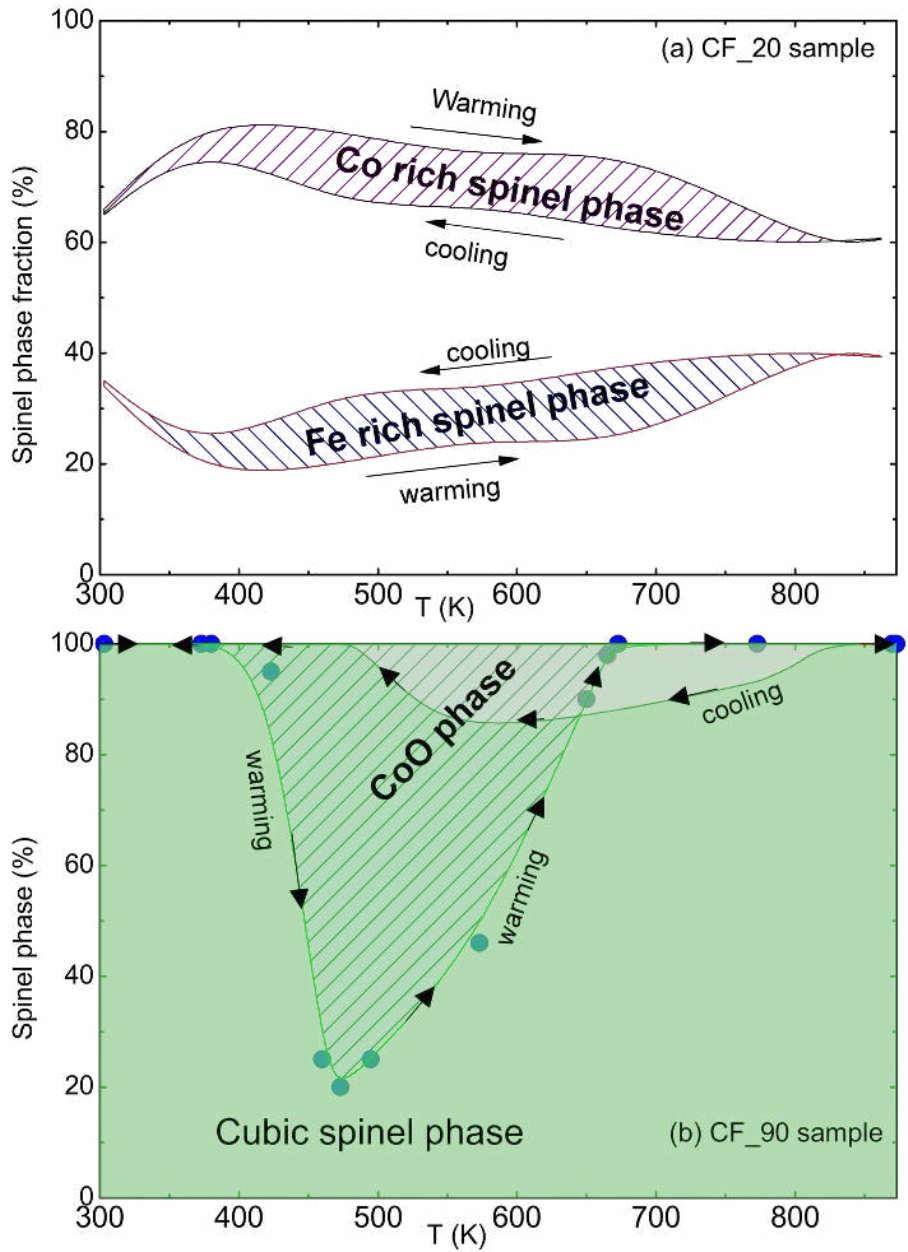

Figure 3 Temperature effect on phase fraction in CF_20 sample (a) and CF_90 sample (b) during warming and cooling modes (guided by Arrows).

parameter during cooling mode suggest an effect of the exchange of Co and Fe ions between A and B sites. Generally, an increasing population of Co ion at B sites reduces lattice parameter in spinel oxides [2-3, 11]. The application of two-phase model shows a large difference between the lattice parameters of Co rich phase (phase 1) and Fe rich phase (phase 2) in CF_20 sample (inset of Figure 4(a)). It may be noted that lattice parameter of the single phase model is close to the

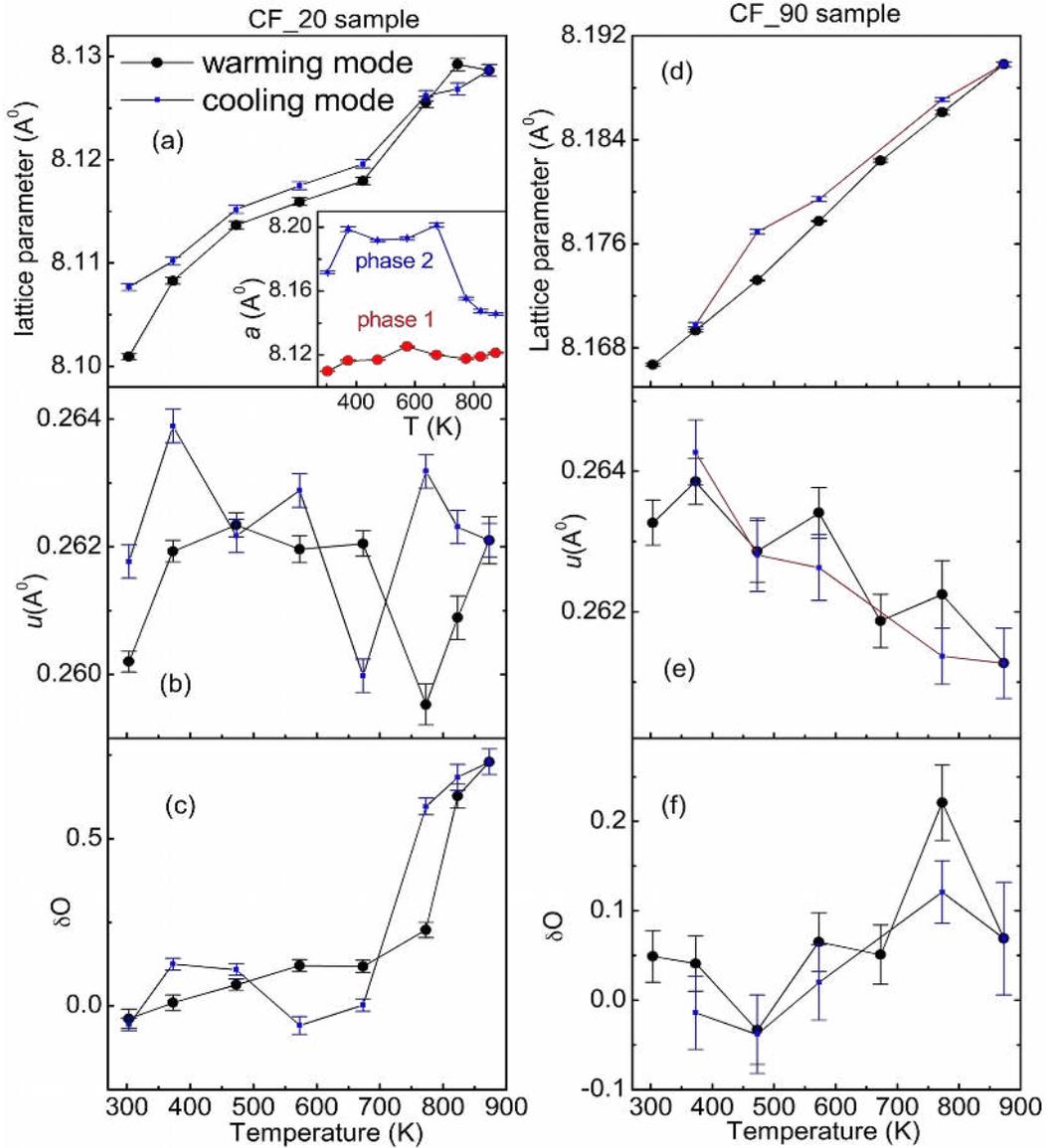

Fiure 4. Structural parameters obtained from Rietveld refinement (single phase) of high temperature SXRD patterns of the CF_20 (a-c) and CF_90 (d-f) samples. The inset of (a) shows lattice parameters for two-phased model of CF_20 sample.

values for Co-rich phase using two-phase model. The parameter, $u$, defines a displacement of O atoms with reference to regular tetrahedron at 8a sites and octahedron at 16d sites. The ideal value of $u$ is 0.25, but usually it lies between 0.24 and 0.275 for spinel oxides [28-29]. Figure 4 (b, e) show a bit scattered values of $u$ (0.259-0.264) for CF_20 sample, whereas a systematic decreasing trend (range: 0.261-0.264) at higher temperatures is found for CF_90 sample. The decreasing trend of $u$ indicates the expansion of B sites at the expense of A sites. This is realized by calculating the M-O (metal-oxygen) bond lengths at A sites ($R_{tet}$) and B sites ($R_{oct}$) using the formula $R_{tet} = a\sqrt{3} \ (u-\frac{1}{8})$ and $R_{oct} = a\sqrt{(3u^2 - 2u + \frac{3}{8})}$ [28] for CF_90 sample. The M-O bond length at A sites decreases by ~ 1.53% (1.963 to 1.933 Å) on increasing the temperature from 300 K to 873 K by increasing the M-O bond length ~ 1.19% (1.936 to 1.959 Å) at B sites. The change of bond length shows chemical disorder (site exchange of Co and Fe ions or the variation of charge state of metal ions) at higher measurement temperatures. In order to understand the temperature variation of ion defects, we have defined the difference of O content ($\delta$O) = $O_{cal}$-$O_{theo}$ per formula unit of the spinel structure, where $O_{cal}$ is the calculated value from refinement using single phase model and $O_{theo}$ is the theoretically assigned value 4. The $\delta$O curves are always positive and a rapid increase at higher temperatures shows absorption of oxygen in spinel structure. The absorption is remarkably high for CF_20 sample. Interestingly, oxygen content regained the normal value of the samples at room temperature after thermal cycling.

The temperature induced structural irreversibility is also realized by analyzing the SXRD peak parameters (intensity, position (2θ) and full width at half maximum (β)), which were determined by Lorentzian shape fit. Figure S1(supplementary) shows intensity of selected peaks at different temperatures, normalized by the peak intensity at 300 K in warming mode. Intensity of the crystalline planes corresponding to spinel structure of CF_20 sample and CF_90 sample reduced during cooling mode in comparison to that during warming mode. Intensity of the planes [(111), (311) and (400)] at lower scattering angles decreased at higher temperatures. The intensity of (111) plane approached to nearly zero at 873 K for CF_20 sample. This peak is highly sensitive to surface defects in spinel oxide [21, 26]. The Intensity of (111) peak is high for CF_90 sample.

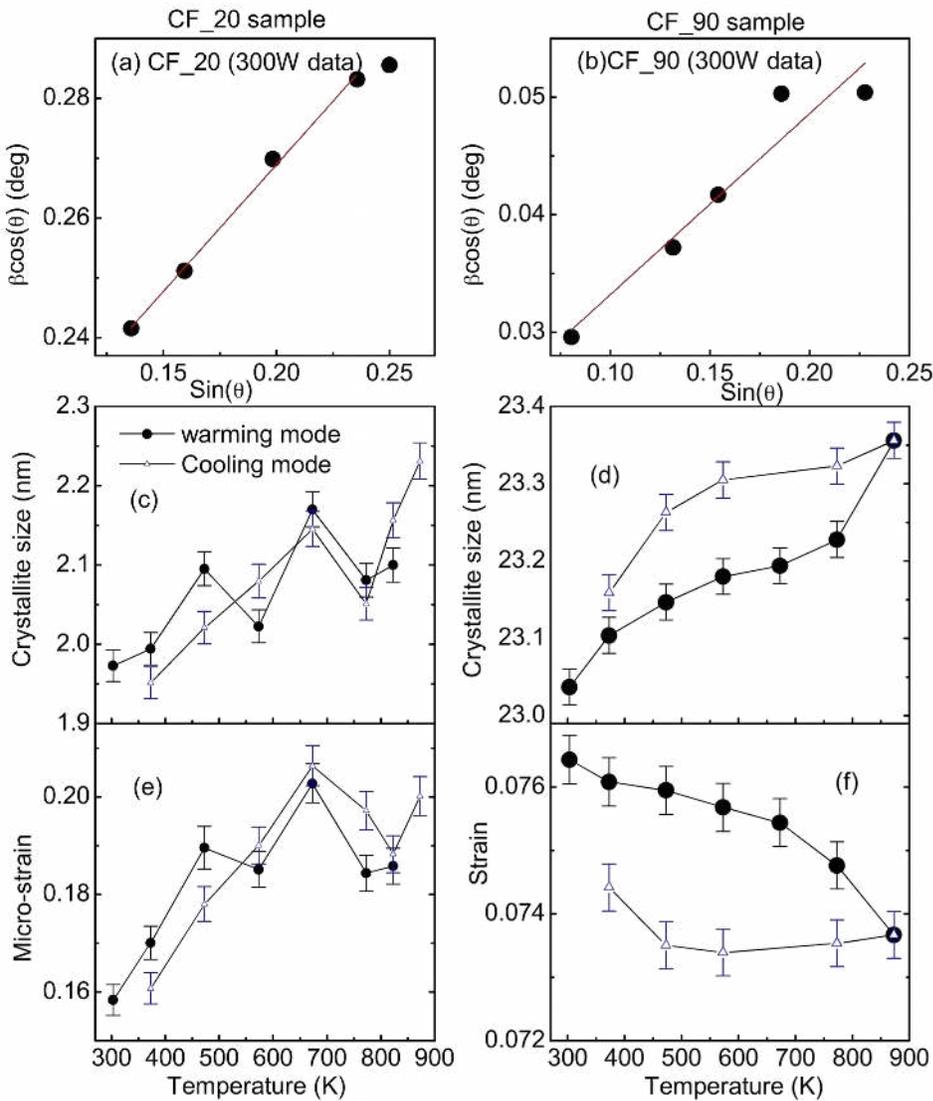

Figure 5. Fit of the SXRD peak parameters at 300 K during warming mode using Williamson-Hall equation (a-b). Temperature dependence of crystallite size (c-d) and micr-strain (e-f) of the samples in warming and cooling modes .

Intensity of the crystalline planes [(422), (333), (440)] at higher scattering angles initially decreases in the temperature range 300 K-473 K, followed by an unusual increase with a broad maximum around 673 K and decreases above 800 K for both the samples during warming mode. Interestingly, intensity of these peaks showed an increasing trend on decreasing the temperature during cooling mode for CF_20 sample, unlike a decreasing trend for CF_90 sample. This shows characteristic differences in irreversibility effect as the function of annealing temperature of the as-prepared material. The crystallite size (<D>) and micro-strain (ε) were calculated using

Williamson-Hall equation: $\beta\cos\theta = \frac{0.89*\lambda}{D} + 2\varepsilon\sin\theta$, where $\lambda$ is wavelength of X-ray radiation.

Figure 5(a-b) shows the fit of $\beta\cos\theta$ vs. $\sin\theta$ data at 300 K (warming mode). The change of crystallite size (Figure 5(c-d)) and micro-strain (Figure 5(e-f)) is considerably small in temperature range 300-873 K, but there is a thermal cycling effect. Crystallite size and micro-strain in CF_20 sample increased with temperature. As an effect of high temperature annealing, CF_90 sample showed larger crystallite size (23.0-23.4 nm) and smaller micro-strain (0.073-0.077) in comparison to smaller crystallite size (1.95-2.25 nm) and larger micro-strain (0.16-0.21) of CF_20 sample.

**Magnetic properties.** Figure 6(a-c) shows the temperature dependence of magnetization curves during field warming (MFW(T)) and field cooling (MFC(T)) modes under magnetic field at 100 Oe or 500 Oe. The MFW(T) curve of CF_50 sample showed superparamagnetic blocking of ferrimagnetic particles/clusters below temperature ($T_m$) at 375 K for applied field 100 Oe [5, 10]. The combined plot of low temperature magnetization and high temperature MFW(T) curves (insets of Figure 6 (a, c)) showed blocking temperature below room temperature for CF_20 sample (~ 270 K for field 100 Oe) and CF_90 sample (~ 230 K for field 500 Oe and at 300 K for field 100 Oe). The MFW(T) curves gradually decreased above the superparamagnetic blocking temperature. Surprisingly, an increment in MFW(T) curves appeared above a typical temperature $T_0^{MFW}$ that depends on annealing temperature of the samples (e.g., 684 K, 690 K and 700 K for the samples CF_20, CF_50 and CF_90, respectively). A local broad maximum is observed at about 880 K for all samples. Height of the magnetization maximum at 880 K dominates over the low temperature maximum at $T_m$ ( $\leq$ 300 K) for CF_90 sample, whereas magnetization maximum at $T_m$ dominates over the high temperature maximum in CF_20 and CF_50 samples.

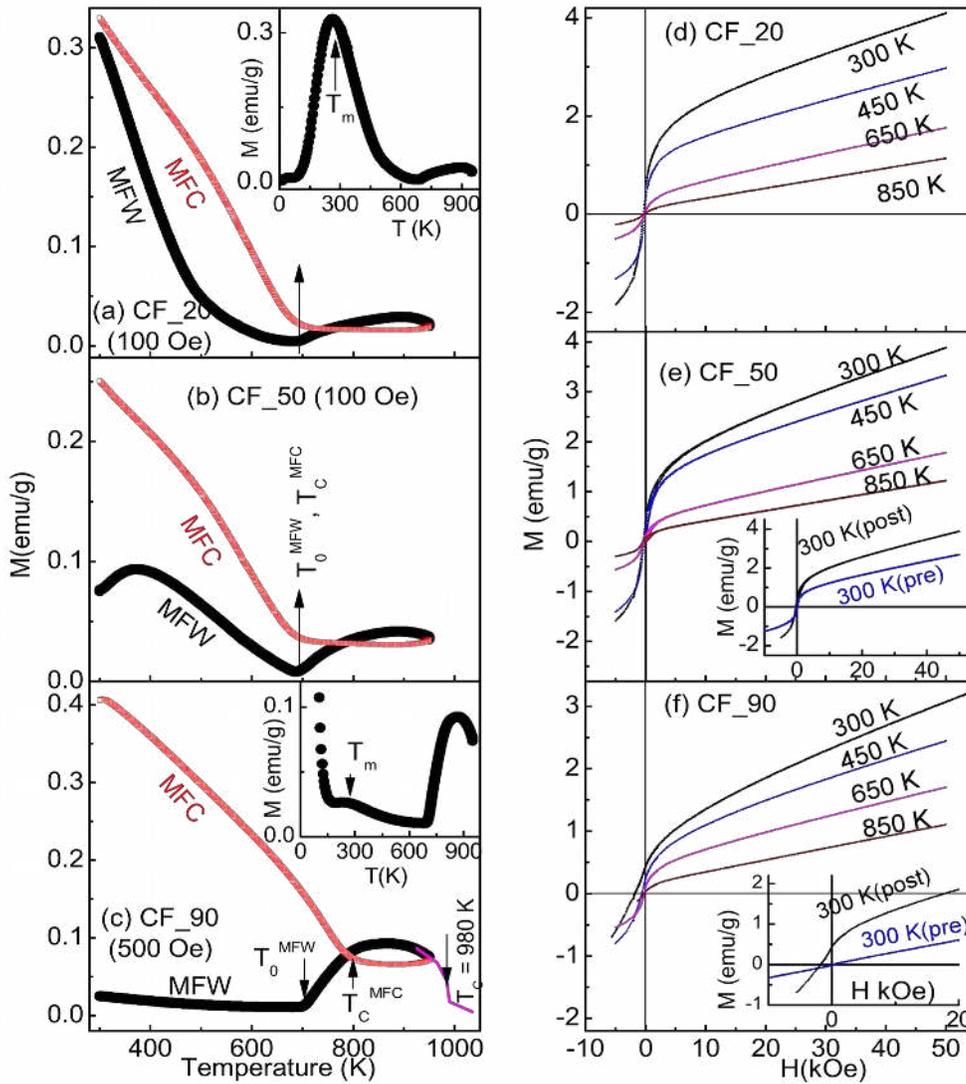

Figure 6. (a-c) The MFW(T) and MFC(T) curves during field warming and field cooling modes. The magnetic transition temperatures are marked by vertical lines. The Insets of (a, c) represent superparamagnetic blocking temperature. (d-f) M(H) curves at selected temperatures. The insets of (e-f) compare M(H) curves at 300 K measured before (pre) and after (post) MFW(T) and MFC(T) measurements.

The unusual properties in MFW(T) curves at higher temperatures is an example of defect induced ferromagnetism, as reported in Fe deficient $Fe_{3-x}Ti_xO_4$ due to $\gamma$-$Fe_2O_3$ surface layer [21]. An extrapolation of the MWF(T) curve for CF_90 sample indicates a $T_C$ at about 980 K, which is in the range of $T_C$ (~ 920 K) for $\gamma$-$Fe_2O_3$ (maghemite) [30]. According to Readman and O'reilly [21], the adsorption and ionization of O atom result in the creation of new sites at A and B sublattices. The created A sites will be filled with $Co^{2+}$ ions, and each new B sites will be filled by $Fe^{3+}/Co^{3+}$ ions. Under oxygen rich environment, a fraction (z) of the $Co^{2+}$ ions at A sites can be converted into $Co^{3+}$ ions through the mechanism $Co^{2+} + \frac{z}{2}O \rightarrow zCo^{3+} + (1-z)\ Co^{2+} + \frac{z}{2}O^{2-}$. A fraction of the $Fe^{3+}/Co^{3+}$ ions at B sites, highly mobile at high temperatures, diffuses into the surface of grains by creating corresponding vacancies at B sites and forms a skin of $\gamma$-$Fe_2O_3$ (maghemite) (or Co doped $\gamma$-$Fe_2O_3$) phase. The interior and skin of the grains rapidly become homogenous within B sites due to migration of vacancies and $Fe^{3+}/Co^{3+}$ ions. At the same time, population of cations at A sites becomes homogeneous by the electrons transfer between interior $Co^{2+}$ ions and surface $Fe^{3+}$ ions. This contributes different character in MFC(T) curves, which followed different paths without any local magnetization maximum during field cooling mode of measurement down to 300 K. The MFC(T) curves showed a sharp increase below a typical temperature $T_C^{MFC}$ (~705 K for CF_20 and CF_50 samples, and 820 K for CF_90 sample). A wide gap between MFC(T) and MFW(T) curves shows substantial change in the magnetic spin order during cooling mode. The M(H) curves (Figure 6 (d-f)) at selected temperatures in the range 300- 900 K and measured in the field range +70/50 kOe to -5 kOe showed ferrimagnetic features with hysteresis loop and lack of magnetic saturation at higher fields. The insets of Figure 6 (e-f) showed enhancement of M(H) curves at 300 K after in-field MFW(T) and MFC(T) measurements (post thermal cycling) than the M(H) curves recorded before high temperature measurement (pre-thermal cycling). The magnetic parameters ($M_S$, $M_R$, $H_C$) of the samples, calculated using M(H) curves, at different temperatures are shown in Figure 7(a-c). A linear fit of the high field M(H) curve on M axis at H = 0 Oe gives the $M_S$ (spontaneous magnetization) value, whereas the intercept of M(H) curve on M axis at H = 0 Oe gives the remanent magnetization ($M_R$) and the intercept of M(H) curve on negative H axis for M = 0 gives the coercivity ($H_C$). The $M_S$(T) values (with units emu/g and $\mu_B$/f.u in Figure 7(a)) decreased at higher temperatures. The rate of decrement became slow above 700 K and there was no unusual peak above 700 K. The $M_S$ values below 600 K are found higher for the samples with low temperature annealing (CF_20 and CF_50) than the CF_90 sample. However, $M_R$ of the CF_20 sample indicted lower values than the values for CF_50 and CF_90 samples (Figure 7(b)). The $H_C$ value increased with annealing temperature of the samples. An increment of the $H_C$ (Figure 7(c)) at higher measurement temperatures may be associated with onset of an additional strain induced magnetic phase above 600 K. In order to understand the nature of magnetic order, we used Arrot plot [25] (see $M^2$ vs. H/M curves in Fig. 7(d-f)). An extrapolation of the linear or polynomial fit of high field $M^2$ vs. H/M curves on positive $M^2$ axis at H/M = 0 corresponds to square of spontaneous magnetization ($M_S^2$(T)). The $M_S$ values from Arrot plot analysis (not shown) are close to the values obtained from direct linear fit of the high field M(H) curves. The $M^2$ vs. H/M plot showed a linear curve at higher fields and a sharp decrease at lower fields with positive slope for all the samples at 300 K. An upward curvature at higher fields and a negative slope at lower fields are observed at temperatures ≥ 350 K/375 K for CF_20 and CF_50 samples

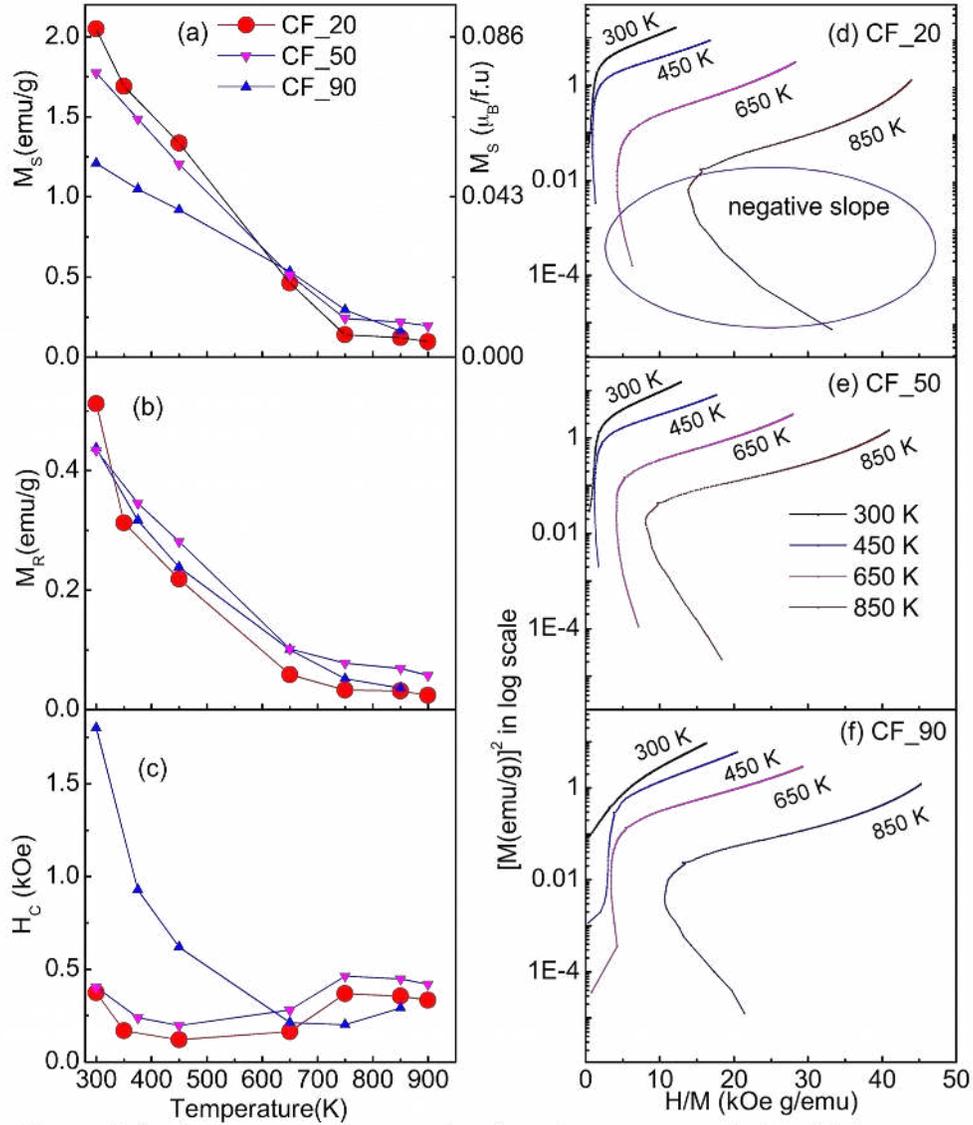

Figgure 7. (a-c) The magnetic parameters (spontaneous magnetization ($M_s$), remanent magnetization ($M_R$) and coercivity ($H_C$)) calculated from M(H) loop of the samples at different temperatures. (d-f) $M^2$ vs. H/M plot of the samples at selected temperatures.

with low annealing temperature, and $\geq 650$ K in case of CF_90 sample. According to Banerjee criterion [31], the slope of H/M vs. $M^2$ plot is positive for second-order transition and negative for first-order phase transition at the boundary between paramagnetic and ferromagnetic phases. The negative slope continuously increased with measurement temperature above 300 K.

**Current-voltage characteristics and Electrical conductivity:** The thermal cycling effect on I-V curve measurement was studied for CF_50 and CF_90 samples during warming (W) and cooling (C) modes. After completing the measurement during first cycle (W1C1), the measurement was repeated during second cycle (W2C2). Figure 8 (a-b) shows I-V characteristics of the CF_50 sample at selected temperatures during W1C1 and W2C2 cycles. The CF_90 sample showed similar I-V characteristics and the data during W2C2 cycle are shown in Figure 8 (c). The I-V characteristics at positive bias voltage are identical to the characteristics at negative bias voltage. We analyzed I-V characteristics at different temperatures using power law ($I = I_0 V^n$) to understand space charge effect. The exponent ($n$) was obtained from the slope of log-log plot of I-V curves. The I-V curves at all temperatures are best fitted with two slopes. The fit and exponent ($n$) values are given in supplementary information (Figure S2). The slope (n1) at low voltage regime is smaller (1-1.2 for CF_50 sample and 1-1.1 for CF_90 sample) in comparison to slope (n2) at high voltage regime (> 10 V). The $n1(T)$ curves showed small irreversibility effect in comparison to $n2(T)$ curves. The higher values (1-1.8) of $n2(T)$ curves show some contribution from interfacial space charges at higher voltage regime [32]. However, $n1$ values close to 1 suggest Ohmic behaviour at low voltage regime. It confirms proper electrical contact between electrode and disc-shaped sample. The resistance (R= V/I) of the samples was calculated from the I-V curves at bias voltages +5 V, +20 V and +40 V. The temperature dependence of resistance (R(T)) curves of the samples showed high resistance (low conductivity) during warming mode and low resistance (high conductivity) during cooling mode. As shown in Figure 8(d-e), the irreversible feature and magnitude of the resistance during W1C1 cycle of the CF_50 sample appeared larger than the W2C2 cycle. By comparing the results in W2C2 cycle, resistance of the CF_50 sample is found smaller (higher conductivity) than that in CF_90 sample. Interestingly, R(T) curves transformed from an unusual metal like state to usual semiconductor state above the temperature $T_{MSW}$ during warming mode and semiconductor to metal like state below the temperature $T_{SMC}$ during cooling mode. The temperature $T_{MSW}$ (~ 443 K) in warming mode of the W1C1 cycle for CF_50 sample is higher than the value (~ 373 K) in W2C2 cycle and the same value (373 K) is found for CF_90 sample. The temperature $T_{SMC}$ is found at about 330 K for both the samples and irrespective of the repetition of cycles. The positive temperature coefficient of resistance may not represent a true metallic state as free electrons are not generally expected in spinel oxides [33-35]. However, trapping and de-trapping of the electronic charges (electrons and holes) in the defect induced localized states at the edges of conduction (CB) and valence (VB) bands can contribute to metal like state in spinel oxide [34]. A detailed discussion has been made later by considering the defects. The R(T) curves in the semiconductor state were fitted with Arrhenius law: $R(T) = R_0 \exp(E_a/k_B T)$. The fit of the lnR vs 1000/T data at 5 V are shown in the insets of Figure 8(a-c). The activation energy ($E_a$) of the CF_50 sample is found to be notably high 3.413 eV during warming mode in comparison to 0.635 eV in the cooling mode of W1C1 cycle. In the W2C2 cycle, activation energy decreased to 0.679 eV and 0.617 eV during warming and cooling modes, respectively. The activation energy of the (high temperature annealed) CF_90 sample in W2C2 cycle was found 0.99 eV and 0.67 eV during warming and cooling modes, respectively. The activation energy in our samples is comparable to the values in spinel oxides [33, 36].

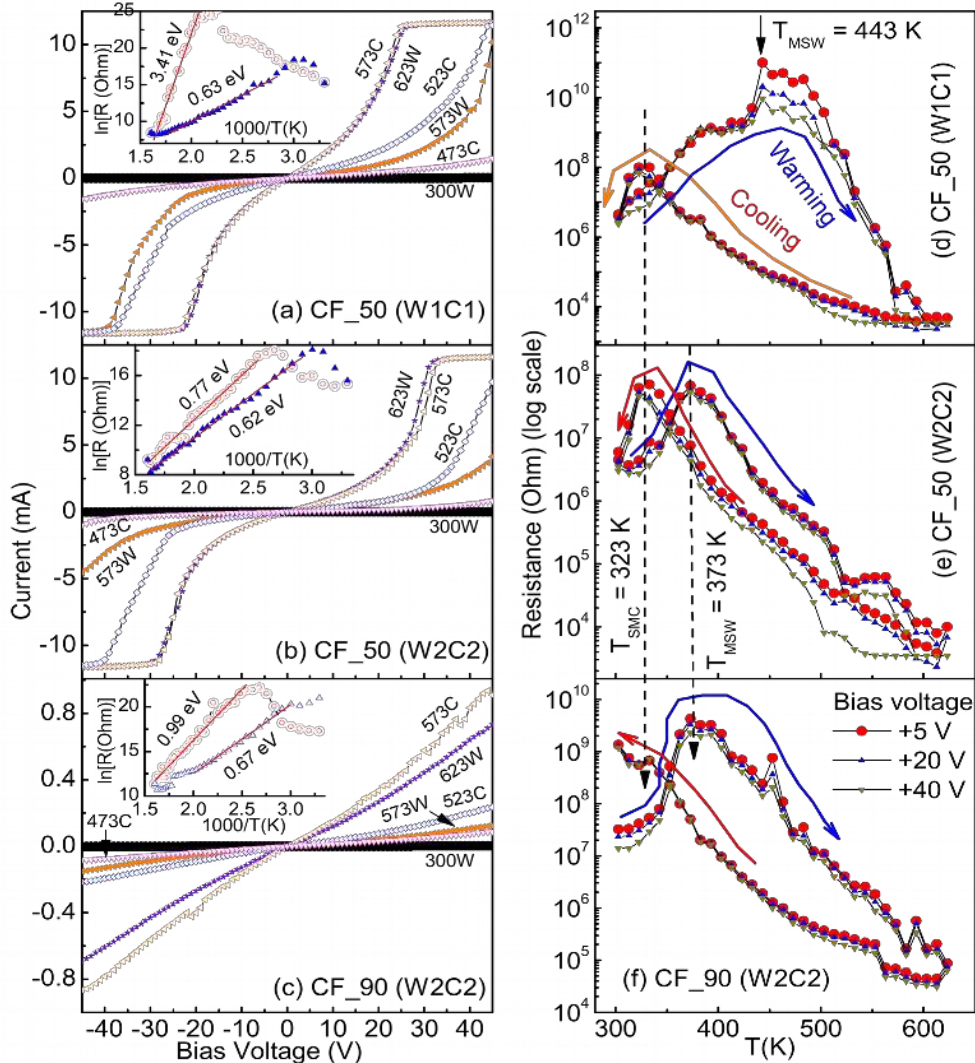

Figure 8. The I-V characteristics at selected temperatures during warming and cooling modes of CF_20 (a-b) and CF_90 (c) samples. The insets show the lnR vs. 1000/T plot at bias voltage +5V. The temperature dependence of resistance of the samples calculated at 5 V, 20 V and 40 V during warming and cooling modes of the measurement (d-f).

■ **DISCUSSION**

Now, we understand a correlation between lattice structure, magnetism and conductivity in the present spinel oxide by referring some of the literature reports. A defect free III−II spinel oxides is supposed to be magnetically ferrimagnet (FIM) and antiferromagnet (AFM), and electrically insulator if all divalent cations ($Fe^{2+}$, $Co^{2+}$) occupy at A sites and all trivalent cations (e.g., $Fe^{3+}$, $Co^{3+}$) occupy B sites. It becomes magnetically disorder and electrically conductive under site exchange of the cations [37]. The semiconductor properties in $Co_xFe_{3-x}O_4$ ferrite can be either *p* type (hole hopping through $Co^{2+}$–$O^{2}$-$Co^{3+}$ superexchange paths for x > 1) or *n* type (electron hopping through $Fe^{3+}$–$O^{2}$-$Fe^{2+}$ superexchange paths for x < 1) [11, 36]. The ideal magnetic spin order and charge conduction are substantially modified in the presence of defects [20]. Huang et al. [22] proposed different intrinsic point defects ($V_{Co(B)}$, $V_{Fe(A)}$, $V_{Fe(B)}$, $V_O$, $Co(B)_{Fe(A)}$, $Co(B)_{Fe(B)}$, $Fe(B)_{Co(B)}$, [$Co(B)_{Fe(A)}$, $Fe(A)_{Co(B)}$]) for $CoFe_2O_4$. The $V_{Co(B)}$ is the vacancy of Co ions at B sites; $V_{Fe(A)}$ is the vacancy of Fe ions at A sites; $Co(B)_{Fe(A)}$ represents the vacancy of Fe ions at A sites substituted by Co ions at B sites, and [$Co(B)_{Fe(A)}$, $Fe(A)_{Co(B)}$] represents exchange between Co ion at B site and Fe ion at A site. The creation of vacancies in spinel structure is site specific. The energy required for creating cation defects at B sites are high in metal (Co/Fe) rich condition and energy values can increase up to 4.72 eV and 3.69 eV for $V_{Fe(B)}$ and $V_{Co(B)}$, respectively. The energy for forming $V_{Fe(A)}$ is high under all conditions and it can go up to 7.20 eV, 5.68 eV and 3.69 eV under metal rich, Co rich and oxygen rich conditions, respectively. The $V_{Co(A)}$ is not energetically favoured, but energy required to form [$Co(B)_{Fe(A)}$, $Fe(A)_{Co(B)}$] through exchange of Fe(A) and Co(B) ions is small (~0.65 eV) and favoured than other defects.

Rietveld refinement confirmed non-stoichiometry in metal/oxygen ratio. The composition of our system is close to $Co_3O_4$ [$(Co^{2+})_A[Co^{3+}Co^{3+}]_BO_4$], where divalent Co ions occupy 1/8 of the (A sites) tetrahedral holes and trivalent ions occupy 1/2 of the (B sites) octahedral holes. Rest of the lattice spaces is vacant and expansion of the lattice space at higher temperature allows adsorption of oxygen atoms. The chemical formula unit can be modified as $A_{1-x1}B_{2-x2}[]_yO_4$ to preserve the oxygen packing in the presence of cation vacancies or $AB_2O_{4+\delta}$ to incorporate excess oxygen [21, 38]. The excess oxygen content ($\delta O$) per formula unit increases at higher measurement temperatures in the present oxide, irrespective of bi-phased or single-phased spinel structure. Of course, non-equilibrium structure of CF_20 sample adsorbed more oxygen atoms and equilibrium spinel structure of CF_90 sample contained small amount of excess oxygen. Considering complex lattice structure of the spinel oxide, certain restrictions in the distribution of Co/Fe ions at A and B sites were followed for simplicity. The structural information may not be affected much by a simple distribution of Co and Fe ions, because both Co and Fe ions have nearly equal number of electrons (atomic scattering factor). It may limit an exact determination of the Co/Fe ratio between A and B sites. However, changes in the Co/Fe ratio between A and B sites can be realized qualitatively from the irreversibility effect in phase fraction, lattice parameters, magnetization, and conductivity difference during thermal cycling [18-19].

Now, we discuss the role of defects on modified magnetism and electrical conductivity. According to Bean-Rodbell theory [39], distorted spin-lattice structure introduces strain induced energy. Net free energy in the system is lowered by distorting the lattice in a direction that increases magnetic exchange energy. A linear relation between the exchange energy constant (λ) and Curie temperature ($T_C$) with the change of cell volume (V) and strength of magnetoelastic coupling (β) can be expressed as λ = $λ_0$(1+$^{\beta \frac{V-V_0}{V_0}}$ ) and $T_C$ = $T_0$(1+$^{\beta \frac{V-V_0}{V_0}}$ ) [25]. The coupling between structural disorder and spin order introduces first order magnetic phase transition [24-25], where free energy corresponding to meta-stable and stable states provides two minima in M(T, H) curves. The MFW(T) curves correspond to meta-stable state (consists of clamped/rigid structure with small lattice distortion at lower temperatures and free structure with strong lattice distortion at higher temperatures) and MFC(T) curves represent a stable state. The FM-PM transition temperature ($T_C^{MFW}$) in MFW(T) curves is expected higher than the PM-FM transition ($T_C^{MFC}$) in MFC(T) curves. We define $T_0^{MFW}$ (~ 680 K-700 K) as the Curie temperature of clamped structure and $T_C^{MFW}$ (~ 980 K) as the Curie temperature of free structure or defect induced ferromagnetic state. In the inter-mediate state between $T_C^{MFW}$ and $T_C^{MFC}$, a thermal hysteresis (MFC(T) <MZFC(T)) is observed during cooling mode because the system remains in PM state until temperature reached close to "true" Curie point ($T_C^{MFC}$) to re-establish its ferro/ ferrimagnetic state. The $T_C^{MFC}$ values are found close to $T_0^{MFW}$ for the samples (CF_20 and CF_50) with low annealing temperature and a substantial difference ($T_C^{MFC} > T_0^{MFW}$) is noted for CF_90 sample. This is due to minor changes in meta-stable state for those samples, where measurement temperature (up to 950 K) is higher than the annealing temperature (473 K for CF_20 sample and 773K for CF_50 sample). In CF_90 sample, meta-stable state is not much affected because measurement temperature (950 K) is less than its annealing temperature 1173 K. The magnetization below 600 K is smaller for CF_90 sample, whose lattice structure is close to $Co_3O_4$ with low magnetic (AFM) state and $T_N \sim$ 30-40 K [40-41], except small replacement of Co by Fe atom. The higher moment in CF_20 and CF_50 samples is contributed by more lattice defects [20, 27]. The local ordering of 3d electronic spins of Co and Fe ions in doubly and triply degenerated $e_g$ and $t_{2g}$ levels of octahedral sites control the magnetic spin order and charge hopping process at B sites of the spinel structure [1]. The magnetic moment in defect free spinel structure is determined by the difference of ionic moments in A and B sites ($M = M_B - M_A$). The $Co^{2+}$ ion (moment ~ 2.52 $\mu_B$) at A sites contains three unpaired electrons at $t_{2g}$ level and at B sites contains one unpaired electron at $t_{2g}$ level and two unpaired electrons at $e_g$ level, respectively. Each $Fe^{3+}$ ion (moment ~ 3.91 $\mu_B$) at A site or B site contains five unpaired electrons at $e_g$ (two) and $t_{2g}$ (three) levels. On the other hand, each $Co^{3+}$ ion at B sites results nearly zero moment due paired six electrons at $t_{2g}$ level. The calculated magnetic moment for $Co^{2+}$ and $Fe^{3+}$ ions in Ref. [1] are relatively smaller than the usual values from spin contribution alone (3 $\mu_B$ for $Co^{2+}$ and 5 $\mu_B$ for $Fe^{3+}$). The smaller value of magnetic moments for $Co^{2+}$ and $Fe^{3+}$ ions in the highly Co rich spinel oxide samples is expected due to the fact of strong hybridization of unoccupied 3d orbital of Fe/Co with 2p orbital of O. The DFT calculations show similar DOS contours for defective and defect-free spinel structure of $CoFe_2O_4$ [22], except extra levels are introduced within band gap in the defective structure. It may be mentioned that an O ion vacancy ($V_O$) produces two electrons (*n* type conductivity) per defect site ($O_O \rightarrow [V_O^0 + 2e'] + \frac{1}{2}O_2(g)$), where as a $Co^{2+}$ ion vacancy ($V_{Co}$) produces two holes (*p* type conductivity) per defect site ($Co_{Co} \rightarrow [V_{Co}^0 + 2h^\cdot] + Co(s)$). In the presence of $V_{Co(B)}^0$ (also $V_{Fe}^0$ both at A and B sites), some occupied levels with spin up ($\uparrow$) states at the Fermi energy and unoccupied levels with spin up states are created in the band gap. In contrast, $V_O^0$ induces extra donor like level occupied by two electrons in the upper part of band gap in spinel oxide. The net moment in spinel structure decreases in the presence of $V_{Fe(B)}$ and $Co(B)_{Fe(B)}$ defects and increases in the presence of $V_{Fe(A)}$ defects. High concentration of defects in Co rich spinel oxide also increases band gap. The exceptionally high activation energy (3.4 eV) during warming mode of the W1C1 cycle measurements for CF_50 sample could be attributed to formation of $V_{Fe(A)}$ defects under high oxygen rich condition. The reduction of its activation energy during W2C2 cycle and also in cooling modes suggests a possibility of different Co/Fe ratio between A and B sites or changes in the fraction of Co-rich/ Fe-rich spinel phases. It may be possible the system slowly approaches to equilibrium structure under repeated thermal cycling, where small activation energy (~0.62-0.99 eV) is needed for the formation of $[Co(B)_{Fe(A)}, Fe(A)_{Co(B)}]$ defects under site exchange of Co and Fe ions [22].

Finally, we suggest that the capacity of adsorbing oxygen atoms during warming mode of high temperature measurement and subsequent release during cooling mode could be promising for studying catalytic properties of surface atoms or surface engineering in Co rich spinel oxide.

This property is highly sensitive to surface response along (111) direction in spinel structure [26] and structural changes in defective spinels during heating cycles [38]. The structure of $Co_3O_4$ spinel oxide along (111) direction has been modeled as a sequence of $Co^{3+}/Co^{2+}$ layers separated by layers of O ions. The surface can by either be A-terminated (by $Co^{2+}$ and $Co^{3+}$ ions) or B-terminated (by $Co^{3+}$ ions) or oxidized on the A (AO) or B (BO) surfaces. In case of the AO and BO terminated surfaces the band gap contains extra states located near to Fermi level. The extra states are partially filled by $d$-orbitals of Co ions and $p$-orbitals of O ions. The number of extra states increases for systems with A- or B-terminated surface. The strongly delocalization of electron states, which almost fill the forbidden gap, gives rise to metal like character and such behavior most probably affected the temperature dependence of resistance curves in our samples. The $Co^{2+}$ ions at A and AO-terminated surfaces have magnetic moment close to that in bulk. The $Co^{3+}$ ions are nonmagnetic in bulk, but they have a magnetic moment close to that of $Co^{2+}$ ions at the B-terminated surface and substantially smaller (non-zero) at the BO surface. The A and AO surfaces contribute AFM properties owing to A-A superexchange interactions, unlike FM properties in B and BO surfaces due to B-B superexchange interactions. Although beyond the scope of the present work, but we believe that a detailed study of high temperature properties can bring out a nice correlation between surface chemistry of metal ions and ferromagnetism [37-40], which is not expected in defect free (conventional) AFM or weak ferrimagnetic spinel oxides.

■ **CONCLUSION**

The structure of $Co_{2.75}Fe_{0.25}O_{4+\delta}$ ($\delta$: 0-0.68) spinel oxide becomes non-stoichiometric during high temperature measurement and re-gains stoichiometric value of oxygen on reversing back to room temperature. The high temperature measurement indicated non-equilibrium lattice structure by showing a preferential orientations of (511) and (440) planes of cubic spinel structure for the sample at low annealing temperature (200 $^0$C) and an additional phase of CoO with preferential (200) plane orientation with thermal hysteresis for the sample annealed at 900 $^0$C. The lattice distortion is more prominent for the samples at low temperature annealing in comparison to the sample annealed at 900 $^0$C. The samples exhibited irreversible effect in the structure, magnetic and electrical properties between warming and cooling modes of the measurements. The meta-stable magnetic state exhibited two magnetic transitions during field warming mode. The defect induced ferromagnetic transition at higher temperature is suppressed during field cooling mode and a ferrimagnetic state is stabilized below a low Curie temperature. The coupling between structural disorder (defects) and magnetic spin order introduces a first order phase transition. We have proposed in graphic abstract a possible correlation between lattice structure, magnetism and electrical conductivity, and promising catalytic application of the studied Co rich spinel oxide.

■ **ASSOCIATED CONTENT**

Additional figures are shown in supplementary information. Table S1 shows the information of Rietveld refinement using SXRD pattern at 873 K and adopting the approach of either variation of Oxygen content at 32e sites or variation of Co content at 16d sites. Figure S1 shows the temperature variation of the SXRD intensity for selected peaks (normalized) of the CF_20 and CF_90 samples during thermal cycling process of SXRD measurements. Figure S2 shows the fit of I-V curves using power law and temperature variation of the power law exponent during thermal cycling process of measurements.

■ **ACKNOWLEDGMENTS**

We acknowledge the Research grant from UGC-DAE- CSR (No M-252/2017/1022), Gov. of India. We thank RRCAT-Indus 2, Indore and UGC-DAE- CSR Mumbai Centre for providing high temperature synchrotron x-ray diffraction and magnetization measurement facilities. RNB thanks Dr. Archana Sagdeo and Mr. M.N. Singh for assisting in SXRD measurements.

# ■ REFERENCES

[1] Walsh, A.; Wei, Su-H.; Yan, Y.; Al-Jassim, M.M.; and Turner, J.A. Structural, magnetic, and electronic properties of the Co-Fe-Al oxide spinel system: Density-functional theory calculations. *Phys. Rev. B* **2007,** *76,* 165119.

[2] Hayashi, K.; Yamada, K.; Shima, M. Compositional dependence of magnetic anisotropy in chemically synthesized $Co_{3-x}Fe_xO_4$ ($0 \leq x \leq 2$). *Jpn. J. Appl.* **2018,** *57*, 01AF02.

[ 3] Liu, S.R.; Ji, D.H.; Xu, J.; Li, Z.Z.; Tang, G.D.; Bian, R.R.; Qi, W.H.; Shang, Z.F.; Zhan,. X.Y. Estimation of cation distribution in spinel ferrites $Co_{1+x}Fe_{2-x}O_4$ ($0.0 \leq x \leq 2.0$) using the magnetic moments measured at 10 K. *J. Alloys Compd.* **2013,** *581*, 616–624.

[4] Bhowmik, R.N.; Panda, M.R.; Yusuf, S.M.; Mukadam, M.D.; Sinha, A.K. Structural phase change in $Co_{2.25}Fe_{0.75}O_4$ spinel oxide by vacuum annealing and role of coexisting CoO phase on magnetic properties. *J. Alloy.Compd.* **2015,** *646*, 161-169.

[5] Panda, M.R.; Bhowmik, R.N.; Singh ,H.; Singh, M.N.; Sinha, A.K. Air annealing effects on lattice structure, charge state distribution of cations, and room temperature ferrimagnetism in the ferrite composition $Co_{2.25}Fe_{0.75}O_4$. *Mater. Res. Exp.* **2015,** *2*, 036101.

[6] Zhang, Y.; Yang, Z.; Zhu ,B.-P.; Ou-Yang, J.; Xiong, R.; Yang, X.-F.; Chen, S. Exchange bias effect of $Co_{3−x}Fe_xO_4$ ($x = 0$, 0.09, 0.14 and 0.27). *J. Alloys Compd.* **2012,** *514*, 25-29.

[7] Bhowmik, R.N.; Kazhugasalamoorthy ,S.; Ranganathan, R.; and Sinha, A.K. Tuning of composite cubic spinel structure in $Co_{1.75}Fe_{1.25}O_4$ spinel oxide by thermal treatment and its effects on modifying the ferromagnetic properties. *J. Alloys Compd.* **2016,** *680*, 315-327.

[8] Bhowmik ,R.N.; Vasanthi, V; and Poddar, A. Alloying of $Fe_3O_4$ and $Co_3O_4$ to develop $Co_{3x}Fe_{3(1-x)}O_4$ ferrite with high magnetic squareness, tunable ferromagnetic parameters, and exchange bias. *J. Alloys Compd.* **2013, 578,** 585-594.

[9] Trong, H.L.; Presmanes, L.; Grave, E.D.; Barnabe, A.; Bonningue, C.; pH Tailhades. Mössbauer characterisations and magnetic properties of iron cobaltites $Co_xFe_{3-x}O_4$ ($1 \leq x \leq 2.46$) before and after spinodal decomposition. *J. Magn. Magn. Mater.* **2013, 334,** 66-73.

[10] Muthuselvam, I.P.; Bhowmik, R.N. Structural phase stability and magnetism in $Co_2FeO_4$ spinel oxide. *Solid State Sci.* **2009, 11,** 719-725.

[11] Bahlawane, N.; Ngamou, P.H.T.; Vannier, V.; Kottke, T.; Heberle, J.; Hoinghaus, K.K. Tailoring the properties and the reactivity of the spinel cobalt oxide. *Phys. Chem. Chem. Phys.* **2009, 11,** 9224-9232.

[12] Debnath, N. et al. As-grown enhancement of spinodal decomposition in spinel cobalt ferrite thin films by Dynamic Aurora pulsed laser deposition, *J. Magn. Magn.Mater.* **2017, 432,** 391-395.

[13] Debnath, N. et al. Magnetic-field-induced phase separation via spinodal decomposition in epitaxial manganese ferrite thin films. Sci. Technol. *Adv. Mater.* **2018, 19,** 507-516.

[14] Bhowmik ,R.N.; Kazhugasalamoorthy, S.; and. Sinha, A.K. Role of initial heat treatment of the ferrite component on magnetic properties in the composite of ferrimagnetic $Co_{1.75}Fe_{1.25}O_4$ ferrite and non-magnetic $BaTiO_3$ oxide. *J. Magn. Magn. Mater.* **2017, 444,** 451-466.

[15] Balagurov, A.M.; Bobrikov, A.; Pomjakushin, V.Yu; Sheptyakov, D.V.; Yushankhai, V. Yu. Interplay between structural and magnetic phase transitions in copper ferrite studied with high-resolution neutron diffraction. *J. Magn. Magn.Mater.* **2015, 374,** 591-599.

[16] Brabers, V.A.M. Infrared Spectra of Cubic and Tetragonal Manganese Ferrites. *Phys. Status Solidi* **1969**, *33*, 563-572.

[17] Nepal, R.; Zhang, Q.; Dai, S.; Tian, W.; Nagler, S.E.; and Jin, R. Structural and magnetic transitions in spinel $FeMn_2O_4$ single crystals. *Phys. Rev. B* **2018**, *97*, 024410.

[18] O'Neill, H.C.; Redfern, S.A.T.; Kesson, S.; and Short, S. An in situ neutron diffraction study of cation disordering in synthetic qandilite $Mg_2TiO_4$ at high temperatures. *Am. Mineral. 2003, 88*, 860-865.

[19] Antao, S.M.; Hassan ,I.; and Parise, J.B. Cation ordering in magnesioferrite, $MgFe_2O_4$, to 982 °C using in situ synchrotron X-ray powder diffraction. *Am. Mineral.* **2005**, *90,* 219-228.

[20] Torres, C.E.R. et al. Oxygen-vacancy-induced local ferromagnetism as a driving mechanism in enhancing the magnetic response of ferrites. *Phys. Rev. B* **2014**, *89,* 104411

[21] Readman, P.W.; and O'Reilly,W. Magnetic Properties of Oxidized (Cation-Deficient) Titanomagnetites (Fe, Ti, ⊣)$_3O_4$. *J. Geomag, Geoelectr.* **1972**, *24*, 69-90.

[22] Huang, Y.L.; Fan, W.B.; Hou, Y.H.; Guo, K.X.; Ouyang, Y.F.; and Liu, Z.W. Effects of intrinsic defects on the electronic structure and magnetic properties of $CoFe_2O_4$: A first-principles study. *J. Magn. Magn.Mater.* **2017**, *429,* 263-269.

[23] Aswathi M C.; and Bhowmik, R.N. Meta-stable magnetic transitions and its field dependence in $Co_{2.75}Fe_{0.25}O_4$ ferrite. *AIP Conf. Proc. 2018,1942*, 130023.

[24] Bustingorry, S.; Pomiro, F.; Aurelio, G.; and Curiale, J. Second-order magnetic critical points at finite magnetic fields: Revisiting Arrott plots. *Phys. Rev. B* **2016**, *93*, 224429.

[25] Nielsen, K.K.; Bahl, C.R.H.; Smith, A.; and Bjørk, R. Spatially resolved modelling of inhomogeneous materials with a first order magnetic phase transition. *J. Phys. D: Appl. Phys.* **2017**, *50,* 414002.

[26] Kupchak, I.; and Serpak, N. Electronic and Magnetic Properties of Spinel $Co_3O_4$ (111) Surface in GGA+U Approximation. *Ukr, J. Phys.* **2017**, *62*, 615**.**

[27] Bhowmik, R. N. Role of interfacial disorder on room temperature ferromagnetism and giant dielectric constant in nano-sized $Co_{1.5}Fe_{1.5}O_4$ ferrite grains. *Mater. Res. Bull.* **2014**, *50,* 476-482.

[28] O'Neill, H.St.C.; and Navrotsky, A. Simple spinels; crystallographic parameters, cation radii, lattice energies, and cation distribution. *Am. Mineral.* **1983**, *68*, 181-194.

[29] Hastings, J.M.; Corliss, L.M. Neutron Diffraction Studies of Zinc Ferrite and Nickel Ferrite, *Rev. Mod. Phys.* **1953**, *25*, 114.

[30] Özdemir, Özden. High-temperature hysteresis and thermoremanence of single-domain maghemite. *Phys. Earth Planet. Inter.* **1990**, *65,* 125-13.

[31] Banerjee, B. K. On a generalised approach to first and second order magnetic transitions.  .*Phys. Lett.* **1964**, *12*, 16-17.

[32] Bhowmik, R.N.; Siva, K.V.; Ranganathan, R.; Mazumdar, C. Doping of Ga in antiferromagnetic semiconductor α-$Cr_2O_3$ and its effects on magnetic and electronic properties. *J. Magn. Magn.Mater.* **2017**, *432*, 56-67.

[33] Bhowmik, R.N.; and Aneesh Kumar, K.S. Role of pH value during material synthesis and grain-grain boundary contribution on the observed semiconductor to metal like conductivity transition in $Ni_{1.5}Fe_{1.5}O_4$ spinel ferrite. *Mater. Chem. Phys.* **2016**, *177*, 417-428.

[34] Younas, M.; Nadeem, M.; Atif, M.; Grossinger, R. Metal-semiconductor transition in $NiFe_2O_4$ nanoparticles due to reverse cationic distribution by impedance spectroscopy. *J. Appl. Phys.* **2011**, *109*, 093704.

[35] Arunkumar, A.; Vanidha, D.; Oudayakumar, K.; Rajagopan, S.; Kannan, R. Metallic magnetism and change of conductivity in the nano to bulk transition of cobalt ferrite. *J. Appl. Phys.* **2013, *114***, 183905.

[36] Jonker, G.H. Analysis of the semiconducting properties of cobalt ferriteJ. Phys. Chem. Solids (1959), 9, (165-175).

 [37] Shi, Y. et al. Self-Doping and Electrical Conductivity in Spinel Oxides: Experimental Validation of Doping Rules. *Chem. Mater.* **2014, *26***, 1867-1873.

[38] Bulavchenko, O.A.; Venediktova, O.S.; Afonasenko, T.N.; Tsyrul'nikov, P.G.; Saraev, A.A.; Kaichev, V.V.;  Tsybulya, S.V. Nonstoichiometric oxygen in Mn–Ga–O spinels: reduction features of the oxides and their catalytic activity. *RSC Adv.* **2018, *8***, 11598-11607.

[39] Bean. C.P.; Rodbell, D.S. Magnetic Disorder as a First-Order Phase Transformation. *Phys. Rev.* **1962, *126***,104.

[40] Ikedo, Y. ; Sugiyama, J.; Nozaki, H.; Itahara, H.; Brewer, J. H.; Ansaldo, E. J.; Morris, G. D.; Andreica, D.; Amato, A. Spatial inhomogeneity of magnetic moments in the cobalt oxide spinel $Co_3O_4$. *Phys. Rev. B* ***2007, 75***, 054424.

[41] Roth, W.L. The magnetic structure of $Co_3O_4$. *J. Phys. Chem. Solids* ***1964, 25***, 1-10.

**Table of contents**

**Synopsis**

Synchrotron XRD, magnetization and current-voltage curves were measured from 300 K to high temperature and back to 300 K for investigation of the irreversibility effect on structure and physical properties. The material was prepared by chemical route and suitably heat treated to form single phase and bi-phased materials. A correlation between lattice structure, magnetism and electrical conductivity, and also a promising catalytic application of the Co rich spinel oxide have been schematically proposed.

**High temperature thermal cycling effect on the irreversible responses of lattice structure, magnetic properties and electrical conductivity in $Co_{2.75}Fe_{0.25}O_{4+\delta}$ spinel oxide**

R.N. Bhowmik[*a], P.D. Babu[b], A.K. Sinha[c,d], and Abhay Bhisikar[c]

[a]*Department of Physics, Pondicherry University, R.V. Nagar, Kalapet, Pondicherry 605014, India*

[b]UGC-DAE Consortium for Scientific Research, Mumbai Centre, Bhabha Atomic Research Centre, Trombay, Mumbai-400085, India

[c]HXAL, SUS, Raja Ramanna Centre for Advanced Technology, Indore- 452013, India

[d]Homi Bhabha National Institute, Anushakti Nagar, Mumbai -400 094 India

[*]Corresponding author: Tel.: +91-9944064547; E-mail: rnbhowmik.phy@pondiuni.edu.in

The cubic spinel structure of $Co_{2.75}Fe_{0.25}O_4$ spinel oxide is modeled close to normal spinel structure $[(Co)_{tet}[Co^{3+}Co^{3+}]_{oct}O_4]$ of $Co_3O_4$ with Wyckoff positions at tetrahedral (8a) sites (1/8, 1/8, 1/8) fully occupied by $Co^{2+}$ ions (or minor occupancy of $Fe^{3+}$ ions), at octahedral (16d) sites (1/2, 1/2, 1/2) co-occupied by $Co^{3+}$ and $Fe^{3+}$ ions, and at 32e sites occupied by oxygen ($O^{2-}$) ions. The lattice structure was refined using single phase and two-phase models. Some of the fit parameters (e.g., isotropic thermal displacements ($B$), oxygen parameter ($u$), occupancy of the Co atoms at tetrahedral sites and Fe at octahedral sites) were suitably refined. The final refinement was carried out by adopting two approaches. In the first approach, occupancy of O atoms at 32e sites was allowed to vary from the value 4 and fixed the 16d sites occupancy of Co atoms and Fe atoms to 1.75 and 0.25, respectively. In the second approach, 16d sites occupancy of Co atoms allowed to vary and fixed occupancy of Fe atoms to 0.25 at 16 sites and occupancy of O atoms to 4 at 32e sites. In case of two phase model for CF_20 sample, we defined Co rich phase as phase 1 (where Co content at 16d sites is assigned more amount than the assigned value for single phase) and the Fe rich phase as phase 2 (where Co content at 16d sites is less and total Fe content at 8a and 16d sites is more than the assigned value for single phase model). The occupancy of Co and Fe atoms at 8a sites and Fe atoms at 16d sites are suitably fixed and final refinement of two phase model was performed either by making occupancy of Co atoms at 16d sites fixed (1.95 for Co rich phase and 1.75 for Fe rich phase) and O atoms variable or occupancy of Co atoms at 16d sites variable and O atoms fixed. We noted that the structural parameters (lattice parameter ($a$), oxygen parameter ($u$)) and the refinement parameters ($R_p$, $R_{wp}$, $R_{exp}$, $\chi^2$) do not differ much whether occupancy of O atoms at 32e site is fixed to 4 and occupancy of Co atoms at 16 sites allowed to vary or occupancy of Co atoms at 16d sites is fixed to 1.75 and occupancy of O atoms at 32e site allowed to vary. Table S1 summarizes structural parameters from refinement of SXRD patterns at 300 K using single phase for CF_90 sample and two-phase model at 873 K for CF_20 sample. Figure S1 shows the temperature variation of the SXRD intensity for selected peaks (normalized) of the CF_20 and CF_90 samples during thermal cycling process of SXRD measurements. The exponent parameters from fit of I-V curves using power law ($I = I_0 V^n$) are given in Figure S2.

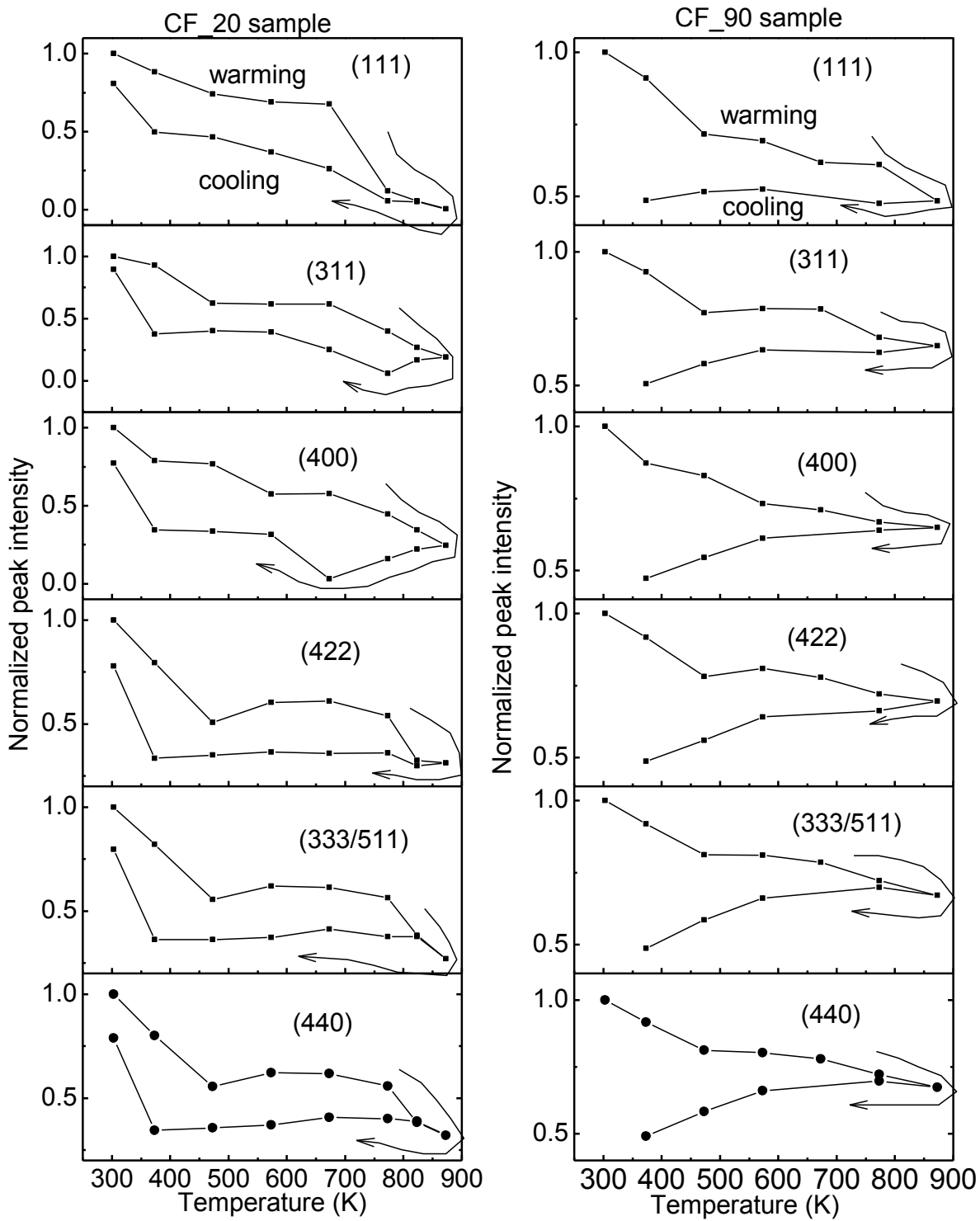

Figure S1 Temperature dependence of SXRD peak intensity in warming and cooling modes for the samples of CF_20 (left side) and CF_90 (right side).

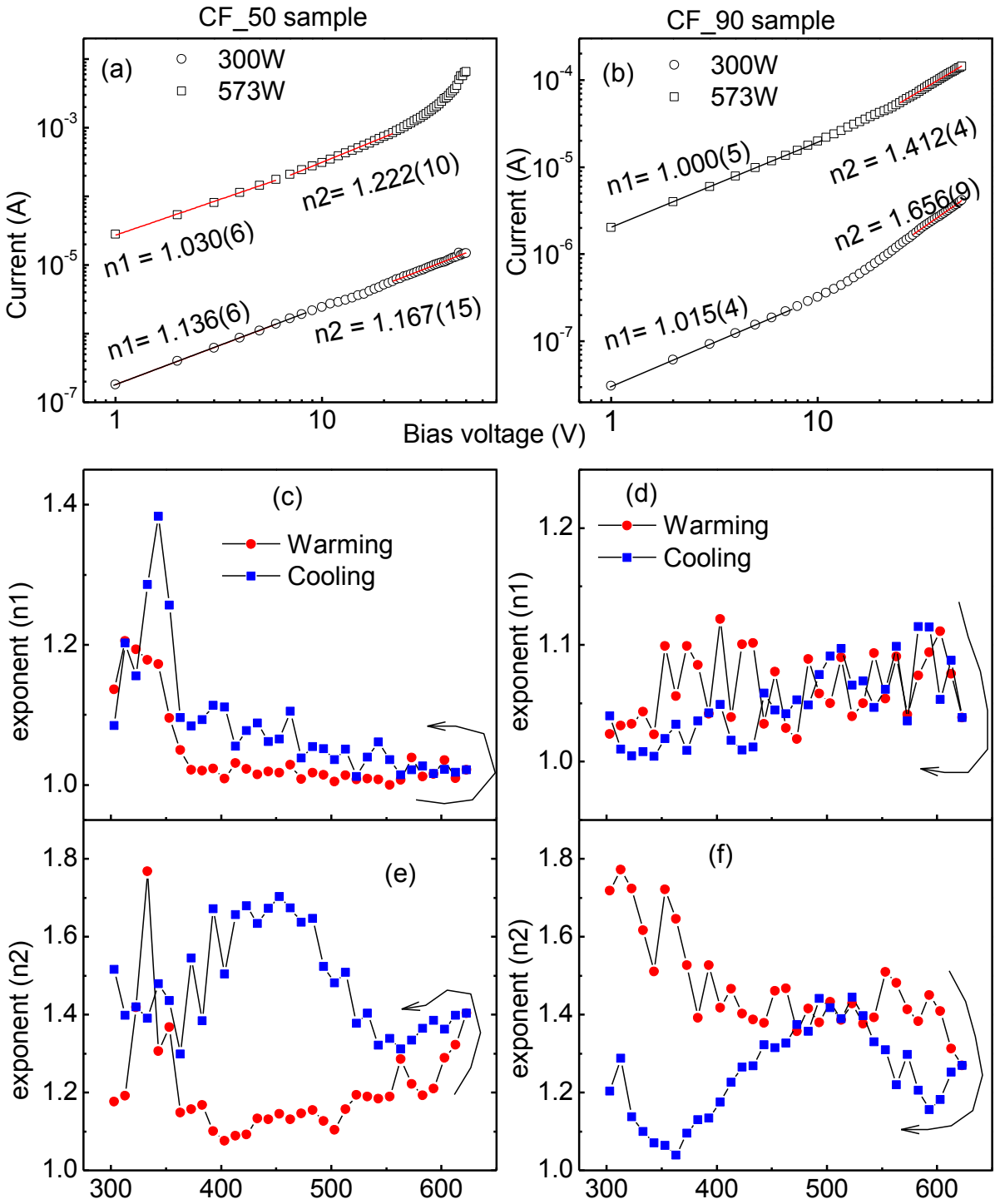

Figure S2. Current vs. voltage curves at 300 K and 573 K in log-log scale to show power law fit (a-b). Temperature variation of the exponent values at low and high voltage regimes for the samples CF_50 (c-d) and CF_90 (e-f), respectively.

Table S1 (supplementary). Rietveld refinement parameters

(a) Refinement of SXRD data at 873 K fitted with single cubic spinel structure for CF_90 sample

| Atoms (sites) | occupancy of O fixed to 4 and Co(16d) variable | | | | | occupancy of Co(16d) fixed to 1.75 and O variable | | | | |
|---|---|---|---|---|---|---|---|---|---|---|
| | Wyckoff positions | | | B | Occupancy | Wyckoff positions | | | B | Occupancy |
| | X | Y | Z | | | X | Y | Z | | |
| **Co(8a)** | 0.12500 | 0.12500 | 0.12500 | 0.97 | 1.000 | 0.12500 | 0.12500 | 0.12500 | 0.97 | 1.000 |
| **Fe(8a)** | 0.12500 | 0.12500 | 0.12500 | 0.97 | 0.000 | 0.12500 | 0.12500 | 0.12500 | 0.97 | 0.000 |
| **Fe(16d)** | 0.50000 | 0.50000 | 0.50000 | 0.97 | 0.250 | 0.50000 | 0.50000 | 0.50000 | 0.97 | 0.250 |
| **Co(16d)** | 0.50000 | 0.50000 | 0.50000 | 0.97 | 1.738(11) | 0.50000 | 0.50000 | 0.50000 | 0.97 | 1.750 |
| **O(32e)** | 0.26127(50) | 0.26127(50) | 0.26127(50) | 0.78 | 4.000(0) | 0.25910(73) | 0.25910(73) | 0.25910(73) | 0.78 | 4.069(64) |

Cell parameter ($a$) = 8.18980(19) Å, volume ($V$) = 549.313(23)Å$^3$    Cell parameter ($a$) = 8.18966 (19) Å, volume ($V$) = 549.285 (22)Å$^3$

$R_p$: 10.5, $R_{wp}$: 13.8, $R_{exp}$: 12.57, $\chi^2$: 1.20    $R_p$: 10.3, $R_{wp}$: 13.7, $R_{exp}$ 12.56, $\chi^2$: 1.19

(b) Refinement of SXRD data at 300 K (warming mode) using two-phase model of cubic spinel structure for CF_20 sample (**O fixed**)

| Atoms (sites) | Co rich phase (Co at 16d sites variable) | | | | | Fe rich phase (Co at 16d sites variable) | | | | |
|---|---|---|---|---|---|---|---|---|---|---|
| | Wyckoff positions | | | B | Occupancy | Wyckoff positions | | | B | Occupancy |
| | X | Y | Z | | | X | Y | Z | | |
| **Co(8a)** | 0.12500 | 0.12500 | 0.12500 | 0.97 | 1.000 | 0.12500 | 0.12500 | 0.12500 | 0.97 | 0.850 |
| **Fe(8a)** | 0.12500 | 0.12500 | 0.12500 | 0.97 | 0.000 | 0.12500 | 0.12500 | 0.12500 | 0.97 | 0.150 |
| **Fe(16d)** | 0.50000 | 0.50000 | 0.50000 | 0.79 | 0.050 | 0.50000 | 0.50000 | 0.50000 | 0.79 | 0.190 |
| **Co(16d)** | 0.50000 | 0.50000 | 0.50000 | 0.79 | 2.026(7) | 0.50000 | 0.50000 | 0.50000 | 0.79 | 1.602(17) |
| **O(32e)** | 0.26047(27) | 0.26047(27) | 0.26047(27) | 0.72 | 4.000 | 0.26212(59) | 0.26212(59) | 0.26212(59) | 0.72 | 4.000 |

Cell parameter (a) = 8.11062 (30) Å, volume (V) = 533.535 (34)Å$^3$    Cell parameter (a) = 8.17313 (73) Å, volume (V) = 545.965 (85)Å$^3$

Co rich phase (Bragg R-factor: 4.77, phase fraction: 64.81%)    $R_p$: 4.12, $R_{wp}$: 5.20, $R_{exp}$: 2.39, $\chi^2$: 4.72

Fe rich phase (Bragg R-factor: 6.21, phase fraction: 35.19%)